\documentclass[trackchanges]{aastex701}

\newcommand{\ca}{\mbox{Ca\,{\textsc{ii}}~K}}
\newcommand{\sunriseiii}{{\sc Sunrise~iii}}

\usepackage{overpic}
\usepackage{xcolor}
\usepackage{url}

\begin{document}

\title{Quantifying the effect of passband on observations in the Ca\,{\sc{ii}}~K~line}

\author[orcid=0009-0003-8344-7544]{Ajay Kumar Yadav}
\affiliation{Max-Planck-Institut für Sonnensystemforschung, Justus-von-Liebig-Weg 3, 37077 Göttingen, Germany}
\email[show]{kumara@mps.mpg.de}  

\author[orcid=0000-0002-0335-9831]{Theodosios Chatzistergos}
\affiliation{Max-Planck-Institut für Sonnensystemforschung, Justus-von-Liebig-Weg 3, 37077 Göttingen, Germany}
\email{chatzistergos@mps.mpg.de}

\author[orcid=0000-0002-1377-3067]{Natalie Krivova} 
\affiliation{Max-Planck-Institut für Sonnensystemforschung, Justus-von-Liebig-Weg 3, 37077 Göttingen, Germany}
\email{natalie@mps.mpg.de}

\author[orcid=0000-0002-3418-8449,sname='Solanki']{Sami~K.~Solanki} \affiliation{Max-Planck-Institut für Sonnensystemforschung, Justus-von-Liebig-Weg 3, 37077 Göttingen, Germany}\email{solanki@mps.mpg.de}

\author[orcid=0000-0003-1409-1145]{Francisco A. Iglesias}
\affiliation{Max-Planck-Institut für Sonnensystemforschung, Justus-von-Liebig-Weg 3, 37077 Göttingen, Germany}
\affiliation{Grupo de Estudios en Heliofísica de Mendoza, CONICET, Universidad de Mendoza, Boulogne sur Mer 683, 5500 Mendoza, Argentina}
\email{iglesias@mps.mpg.de}

\author[0000-0003-2596-9523]{Ilaria~Ermolli}
\affiliation{INAF Osservatorio Astronomico di Roma, Via Frascati 33, 00078 Monte Porzio Catone, Italy}\email{ilaria.ermollio@inaf.it}

\author[orcid=0000-0003-1459-7074,sname='Lagg']{Andreas~Lagg} \affiliation{Max-Planck-Institut für Sonnensystemforschung, Justus-von-Liebig-Weg 3, 37077 Göttingen, Germany}\email{lagg@mps.mpg.de}

\author[orcid=0000-0002-9972-9840,sname='Gandorfer']{Achim~Gandorfer} \affiliation{Max-Planck-Institut für Sonnensystemforschung, Justus-von-Liebig-Weg 3, 37077 Göttingen, Germany}\email{gandorfer@mps.mpg.de}

\author[orcid=0000-0002-3387-026X,sname='del~Toro~Iniesta']{Jose~Carlos~del~Toro~Iniesta} \affiliation{Instituto de Astrofísica de Andalucía, CSIC, Glorieta de la Astronomía s/n, 18008 Granada, Spain}\affiliation{Spanish Space Solar Physics Consortium}\email{jti@iaa.es}	

\author[orcid=0000-0002-5054-8782,sname='Katsukawa']{Yukio~Katsukawa} \affiliation{National Astronomical Observatory of Japan, 2-21-1 Osawa, Mitaka, Tokyo 181-8588, Japan}\affiliation{Department of Earth and Planetary Science, The University of Tokyo, 7-3-1, Hongo, Bunkyo-ku, Tokyo 113-0033, Japan}\affiliation{Department of Astronomical Science, The Graduate University for Advanced Studies (SOKENDAI), 2-21-1 Osawa, Mitaka, Tokyo 1818588, Japan}\email{yukio.katsukawa@nao.ac.jp}	

\author[orcid=0000-0002-0787-8954,sname='Bernasconi']{Pietro~Bernasconi} \affiliation{Johns Hopkins University Applied Physics Laboratory, 11100 Johns Hopkins Road, Laurel, Maryland, USA}\email{pietro.bernasconi@jhuapl.edu}

\author[sname='Berkefeld']{Thomas~Berkefeld} \affiliation{Institut für Sonnenphysik (KIS), Georges-Köhler-Allee 401a, 79110 Freiburg, Germany}\email{thomas.berkefeld@leibniz-kis.de}	

\author[orcid=0009-0009-4425-599X,sname='Feller']{Alex~Feller} \affiliation{Max-Planck-Institut für Sonnensystemforschung, Justus-von-Liebig-Weg 3, 37077 Göttingen, Germany}\email{feller@mps.mpg.de}	

\author[orcid=0000-0001-6317-4380,sname='Riethmüller']{Tino~L.~Riethmüller} \affiliation{Max-Planck-Institut für Sonnensystemforschung, Justus-von-Liebig-Weg 3, 37077 Göttingen, Germany}\email{riethmueller@mps.mpg.de}	

\author[orcid=0000-0001-9228-3412,sname='Álvarez-Herrero']{Alberto~Álvarez-Herrero} \affiliation{Instituto Nacional de T\'ecnica Aeroespacial (INTA), Ctra. de Ajalvir, km. 4, E-28850 Torrejón de Ardoz, Spain}\affiliation{Spanish Space Solar Physics Consortium}\email{alvareza@inta.es}		

\author[orcid=0000-0001-5616-2808,sname='Kubo']{Masahito~Kubo} \affiliation{National Astronomical Observatory of Japan, 2-21-1 Osawa, Mitaka, Tokyo 181-8588, Japan}\email{masahito.kubo@nao.ac.jp}		

\author[orcid=0000-0003-3490-6532,sname='Smitha']{H.~N.~Smitha} \affiliation{Max-Planck-Institut für Sonnensystemforschung, Justus-von-Liebig-Weg 3, 37077 Göttingen, Germany}\email{narayanamurthy@mps.mpg.de}	

\author[orcid=0000-0001-8829-1938,sname='Orozco~Suárez']{David~Orozco~Suárez} \affiliation{Instituto de Astrofísica de Andalucía, CSIC, Glorieta de la Astronomía s/n, 18008 Granada, Spain}\affiliation{Spanish Space Solar Physics Consortium}\email{orozco@iaa.es}	

\author[sname='Grauf']{Bianca~Grauf} \affiliation{Max-Planck-Institut für Sonnensystemforschung, Justus-von-Liebig-Weg 3, 37077 Göttingen, Germany}\email{grauf@mps.mpg.de}		

\author[sname='Carpenter']{Michael~Carpenter} \affiliation{Johns Hopkins University Applied Physics Laboratory, 11100 Johns Hopkins Road, Laurel, Maryland, USA}\email{michael.carpenter@jhuapl.edu}

\author[sname='Bell']{Alexander~Bell} \affiliation{Institut für Sonnenphysik (KIS), Georges-Köhler-Allee 401a, 79110 Freiburg, Germany}\email{albe@leibniz-kis.de}

\author[orcid=0000-0001-7764-6895, sname={Martínez~Pillet}]{Valentín~Martínez~Pillet}
\affiliation{Instituto de Astrofísica de Canarias, Vía Láctea, s/n, E-38205 La Laguna, Spain}
\affiliation{Spanish Space Solar Physics Consortium}
\email{vmpillet@iac.es}

\author[orcid=0000-0001-7696-8665,sname='Gizon']{Laurent~Gizon} \affiliation{Max-Planck-Institut für Sonnensystemforschung, Justus-von-Liebig-Weg 3, 37077 Göttingen, Germany}\affiliation{Institut für Astrophysik und Geophysik, Georg-August-Universität Göttingen, 37077 Göttingen, Germany}\email{gizon@mps.mpg.de}

\author[orcid=0000-0001-6029-7529,sname='Hoelken']{Johannes~Hoelken} \affiliation{Max-Planck-Institut für Sonnensystemforschung, Justus-von-Liebig-Weg 3, 37077 Göttingen, Germany}\email{hoelken@mps.mpg.de}	

\author[orcid=0000-0002-7318-3536, sname={Bailén}]{Francisco~Javier~Bailén}
\affiliation{Instituto de Astrofísica de Andalucía, CSIC, Glorieta de la Astronomía s/n, 18008 Granada, Spain}
\affiliation{Spanish Space Solar Physics Consortium}
\email{fbailen@iaa.es}

\author[orcid=0000-0002-2055-441X,sname='Blanco~Rodríguez']{Julian~Blanco~Rodríguez} \affiliation{Universitat de Valencia Catedrático José Beltrán 2, E-46980 Paterna-Valencia, Spain}\affiliation{Spanish Space Solar Physics Consortium}\email{julian.blanco@uv.es}	

\author[orcid=0000-0003-4319-2009,sname='Castellanos~Durán']{Juan~Sebastián~Castellanos~Durán} \affiliation{Max-Planck-Institut für Sonnensystemforschung, Justus-von-Liebig-Weg 3, 37077 Göttingen, Germany}\email{castellanos@mps.mpg.de}		

\author[orcid=0009-0002-6808-5154,sname='Harnes']{Edvarda~Harnes} \affiliation{Max-Planck-Institut für Sonnensystemforschung, Justus-von-Liebig-Weg 3, 37077 Göttingen, Germany}\email{harnes@mps.mpg.de}

\author[orcid=0000-0002-4669-5376,sname='Ishikawa']{Ryohtaroh~T.~Ishikawa} \affiliation{National Institute for Fusion Science, 322-6 Oroshi-cho, Toki City 509-5292, Japan}\email{ishikawa.ryohtaro@nifs.ac.jp}		

\author[orcid=0000-0001-7452-0656,sname='Kawabata']{Yusuke~Kawabata} \affiliation{National Astronomical Observatory of Japan, 2-21-1 Osawa, Mitaka, Tokyo 181-8588, Japan}\email{kawabata.yusuke@nao.ac.jp}		

\author[orcid=0000-0002-1043-9944,sname='Matsumoto']{Takuma~Matsumoto} \affiliation{Centre for Integrated Data Science, Institute for Space-Earth Environmental Research, Nagoya University, Furocho, Chikusa-ku, Nagoya, Aichi 464-8601, Japan}\email{takuma.matsumoto@gmail.com}		

\author[orcid=0000-0002-7044-6281,sname='Oba']{Takayoshi~Oba} \affiliation{Advanced Research Center for Space Science and Technology, Institute of Science and Engineering, Kanazawa University, Kakuma-machi, Kanazawa, Ishikawa 920-1192, Japan}\affiliation{Max-Planck-Institut für Sonnensystemforschung, Justus-von-Liebig-Weg 3, 37077 Göttingen, Germany}\email{n_mus_gi@yahoo.co.jp}		

\author[orcid=0000-0003-0175-6232,sname='Siu-Tapia']{Azaymi~L.~Siu-Tapia} \affiliation{Instituto de Astrofísica de Andalucía, CSIC, Glorieta de la Astronomía s/n, 18008 Granada, Spain}\affiliation{Spanish Space Solar Physics Consortium}\email{siu@iaa.es}		

\author[orcid=0000-0003-1483-4535,sname='Strecker']{Hanna~Strecker} \affiliation{Instituto de Astrofísica de Andalucía, CSIC, Glorieta de la Astronomía s/n, 18008 Granada, Spain}\affiliation{Spanish Space Solar Physics Consortium}\email{streckerh@iaa.es}		

\author[orcid=0000-0003-1971-5551,sname='Vukadinović']{Dušan~Vukadinović} \affiliation{Institut für Physik, Universität Graz, Universitätsplatz 5, 8010 Graz, Austria}\affiliation{Max-Planck-Institut für Sonnensystemforschung, Justus-von-Liebig-Weg 3, 37077 Göttingen, Germany}\email{vukadinovic@mps.mpg.de}	

\author[orcid=0000-0002-5332-8881,sname='Narita']{Yasuhito Narita}
\affiliation{ Institut für Theoretische Physik, Technische Universität Braunschweig, Mendelssohnstr. 3, 38106 Braunschweig, Germany}\affiliation{Max-Planck-Institut für Sonnensystemforschung, Justus-von-Liebig-Weg 3, 37077 Göttingen, Germany}\email{y.narita@tu-braunschweig.de}

\begin{abstract}
Full-disk observations of the Sun in the \ca\ line have been carried out since the late 19th century at various observatories worldwide. 
These long-term records of solar activity are crucial for reducing discrepancies among solar irradiance reconstructions and for advancing our understanding of the solar dynamo.
To construct a consistent composite record, data from different observatories must be cross-calibrated to account for variations in spectral passband and spatial resolution, which are the primary sources of discrepancies between archives. 
In this study, we use high spectral and spatial resolution observations in the \ca\ line from the state-of-the-art \sunriseiii\ mission to emulate different passbands and derive empirical contrast-contrast relationships between them. We find that these relationships are well described by a power law and provide coefficients for different combinations of passband widths in the range 0.1--9\,\AA\ and spatial resolutions between 1\arcsec\ and 6\arcsec. Applying such a relationship to observations from two major \ca\ archives demonstrates its potential to improve their cross-calibration.
The results provide a foundation for the construction of a consistent,  century-long time series of solar activity from historical and modern \ca\ observations.
\end{abstract}


\section{Introduction}
Solar radiation is the primary energy source for Earth's system \citep[e.g.,][]{kren2017}. 
This energy~-- quantified by solar irradiance, which is the wavelength-dependent power received per unit area outside Earth's atmosphere at one~AU~-- has been monitored for the last five decades \citep[see review by][]{kopp_solar_2025} and found to vary on all observable timescales \citep{Frohlich2006}. 
The variability is primarily driven by the evolution of solar surface magnetism
\citep{Shapiro_2017, Yeo2017}.
Long-term changes in irradiance may influence Earth's climate \citep{Haigh2007,Gray2010,Solanki2013}, but the short span of direct measurements requires extension of the irradiance record into the pre-satellite era with the help of models.

Models ranging from simple regression-based to more sophisticated semi-empirical physics-based approaches have been used for irradiance reconstruction \citep[see reviews by][]{Solanki2013, Ermolli2013,chatzistergos_long-term_2023}. 
These models successfully reproduce observed irradiance variability \citep[e.g.,][]{krivova2003, Yeo2017,chatzistergos_revisiting_2025}, 
but all require information on the evolution of sunspots and faculae.
While such data exist for the satellite era in the form of full-disk magnetograms, reliable and reproducible facular observations are scarce before that period, forcing models to infer surface facular coverage indirectly from sunspot indices. This introduces major uncertainties and contributes to the spread among long-term irradiance reconstructions and
in their estimates of the secular variability \citep[for a review see][]{chatzistergos_long-term_2023}.

A unique opportunity to extend direct facular information comes from
observations in the \ca\ line,  carried out at various observatories worldwide since the late 19th century \citep[see review by][]{chatzistergos2022}.
\ca\ observations clearly show plage regions, which are the chromospheric counterparts of faculae, making them a valuable proxy of magnetic activity.
If properly calibrated, these historical and modern \ca\ archives  could provide direct facular information and therefore greatly reduce uncertainties in long-term irradiance reconstruction. 

Furthermore, \ca\ brightness has been shown to correlate tightly  with the photospheric magnetic field \citep{babcock1955,Loukitcheva,Schrijver}, which also enables the reconstruction of pseudo-magnetograms from \ca\ images \citep[e.g.][]{Pevtsov, Chatzistergos2019_ca_to_b}. 
Hence, combining historical and modern \ca\ observations could yield a consistent, century-long record of solar (unsigned) magnetograms. These magnetograms can then be used for irradiance reconstruction \citep{Dasi_2014, Dasi_2016}, understanding the long-term evolution of the photospheric magnetic field \citep{Hofer_2024}, and solar dynamo studies \citep{Jin_and_wang}.  

Despite this potential, the use  of \ca\ data has long been limited due to numerous issues with their processing.
Substantial progress has been achieved in recent years \citep[e.g.][]{Chatzistergos_analysis_2019,chatzistergos_analysis_2020,chatzistergos2021,chatzistergos_understanding_2024}, although some significant challenges remain.
These include, for example, understanding the effect of instrumental changes and degradations on the data. 
More critically, the various \ca\ archives were obtained  with different instrumental setups with distinct spectral passbands and spatial resolutions. 
As a consequence, images from different observatories sample different atmospheric heights and spatial scales. 
As a result, most previous studies either were limited to a single archive or combined multiple archives without properly accounting for their differences, thereby risking inhomogeneities and spurious trends \citep[see][and references therein]{Chatzistergos2019_ca_to_b}.
Exceptions were the studies by \citet{chatzistergos2021,chatzistergos_revisiting_2025}, who normalized images from different observatories by the standard deviation of the quiet-Sun (QS) regions, but
noted that the normalization factor might be non-linear. 

Quantifying the influence of instrumental passband and spatial resolution on \ca\ brightness is therefore essential for reliable cross-calibration and homogenization of historical and modern datasets.
To this end, \citet{murabito_investigating_2023} analyzed \ca\ observations from the Swedish Solar Telescope, degrading the spectral and spatial resolution to match those of historical instruments in order to assess how these factors affect the observed signal and line profiles. While their work improved our understanding of the impact of these factors on specific line parameters, it was limited to regional averages. 
A practical framework enabling the direct, pixel-by-pixel conversion of full-disk solar \ca\ images between different instrumental configurations has so far been lacking, but is needed to produce a homogeneous time series of \ca\ observations and of any parameters deduced therefrom.

Here, we take the next step by systematically evaluating how passband width influences the observed brightness and, crucially, by deriving empirical pixel-by-pixel relationships that allow observations obtained with a given passband and spatial resolution to be translated to different instrumental setups. We overcome previous limitations for such a study by employing state-of-the-art data from the Sunrise UV Spectropolarimeter and Imager (SUSI) taken during the 2024 flight of the \sunriseiii\ balloon-borne observatory \citep{Lagg_2025,solanki2026}, which provided spectra around the \ca\  line with spectral sampling of $10$~m\AA~\text{pixel}$^{-1}$ \citep{Feller2025}. \sunriseiii\ builds upon the original {\sc Sunrise} observatory \citep{Barthol_2011}, which flew successfully in 2009 and 2013 \citep{Solanki_2010, Solanki_2017}.

\section{Data}
\label{sec:data}

To derive the relationships between \ca\ brightness as measured through different passbands, we use observations made with the ultraviolet slit spectropolarimeter \mbox{\sunriseiii/SUSI}.
The dataset comprises one raster scan of a plage region (32\_PLAG in Table 1 of \citealt{solanki2026}, referred to as Plage hereafter,  pointing at x=-125\arcsec, y=189\arcsec\ taken on 15 July 2024 starting from 17:43\,UT lasting 55 minutes) and two consecutive scans of an emerging flux region (02\_EMEF in \citealt{solanki2026}, called EMEF~1 and EMEF~2, hereafter, pointing at x=-71\arcsec, y=-199\arcsec\ obtained on 10 July 2024 at 19:14 UT each 37 minutes long), see panels a--c of Figure~\ref{fig:emef_1_2_plage_cont_2_5_passband_spot_masked} and figure in Appendix \ref{appendix:qs_region}. All observations were acquired at $\mu=0.97$ (where $\mu$ is the cosine of the heliocentric angle), with an effective integration time per slit position of 3.3\,s, and spatial and spectral samplings of 0.03\arcsec~$\text{pixel}^{-1}$ and  $10$~m\AA~\text{pixel}$^{-1}$, respectively.

SUSI data were reduced with the standard spectropolarimetric pipeline using both pre-flight and in-flight calibration data \citep{Feller2025, solanki2026}. The reduction includes corrections for dark field, variable camera bias, static optical distortions, and flat field. Spectral dispersion and continuum level calibrations were performed by comparing spatially-averaged disk-center SUSI data with the Hamburg Fourier transform spectrograph atlas (FTS, \citealt{Neckel1999}), see \cite{Hoelken2024} for details. 
The polarimetric data were demodulated to obtain the full Stokes vector, although in this work we employ only the intensity spectra, see \cite{Iglesias2025} for details. 
No numeric image restoration or spectral deconvolution  was applied.

To test the relationships derived from the \sunriseiii\ data, we applied them to filtergram observations in the \ca\ line from two ground-based observatories: Rome/PSPT \citep[Precision Solar Photometric Telescope;][]{ermolli_rome_2022} and the Meudon heliograph\footnote{Available at \url{ftp://ftpbass2000.obspm.fr/2012/meudon/CaIIk/1208/}}, which were recorded on  1 August 2012 at 07:08:37~UT and 07:24:54~UT, respectively.
The Rome filtergrams have a nominal passband of 2.5~\AA\ and a pixel scale of 2\arcsec~$\text{pixel}^{-1}$,
while Meudon uses a 1.4~\AA\ passband and 1\arcsec~$\text{pixel}^{-1}$ sampling; both are centered at the core of \ca\ line.

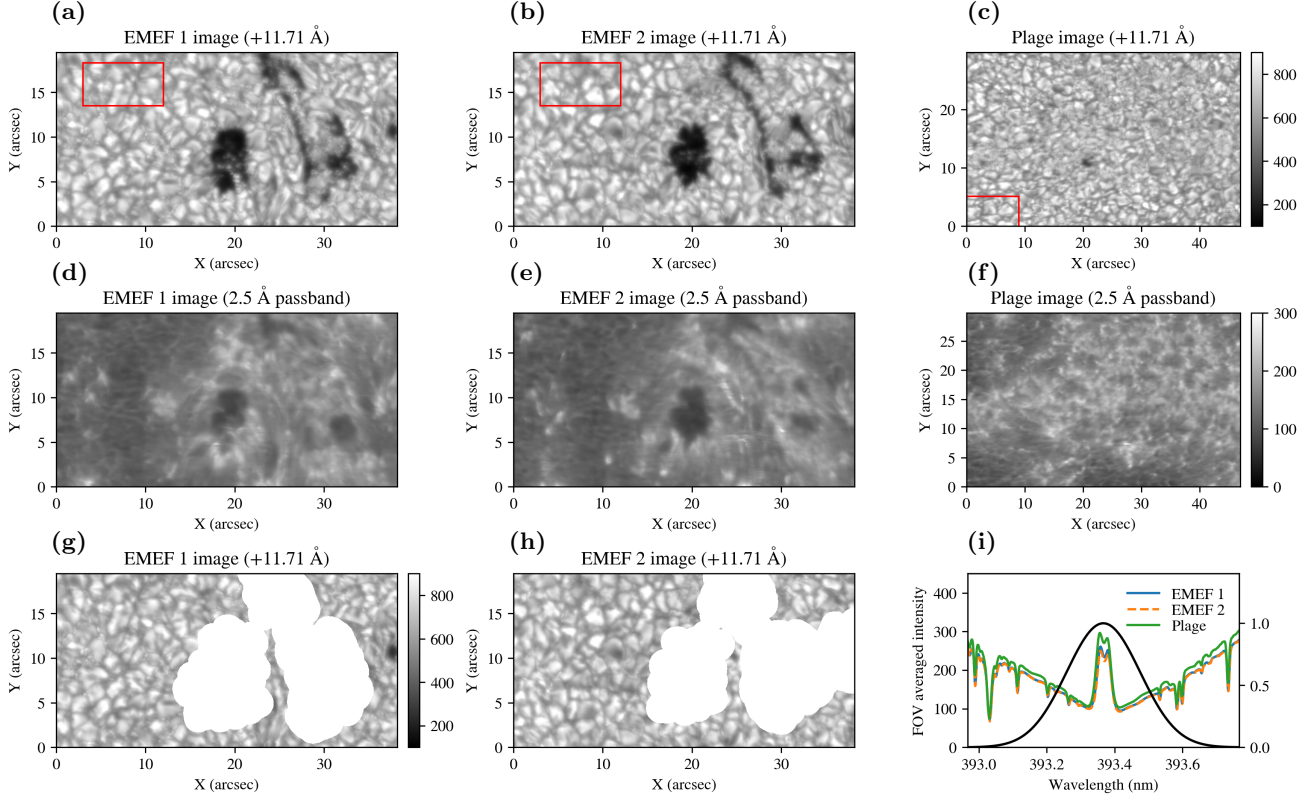
\begin{figure}
    \centering
    \begin{overpic}[width=0.96\linewidth]{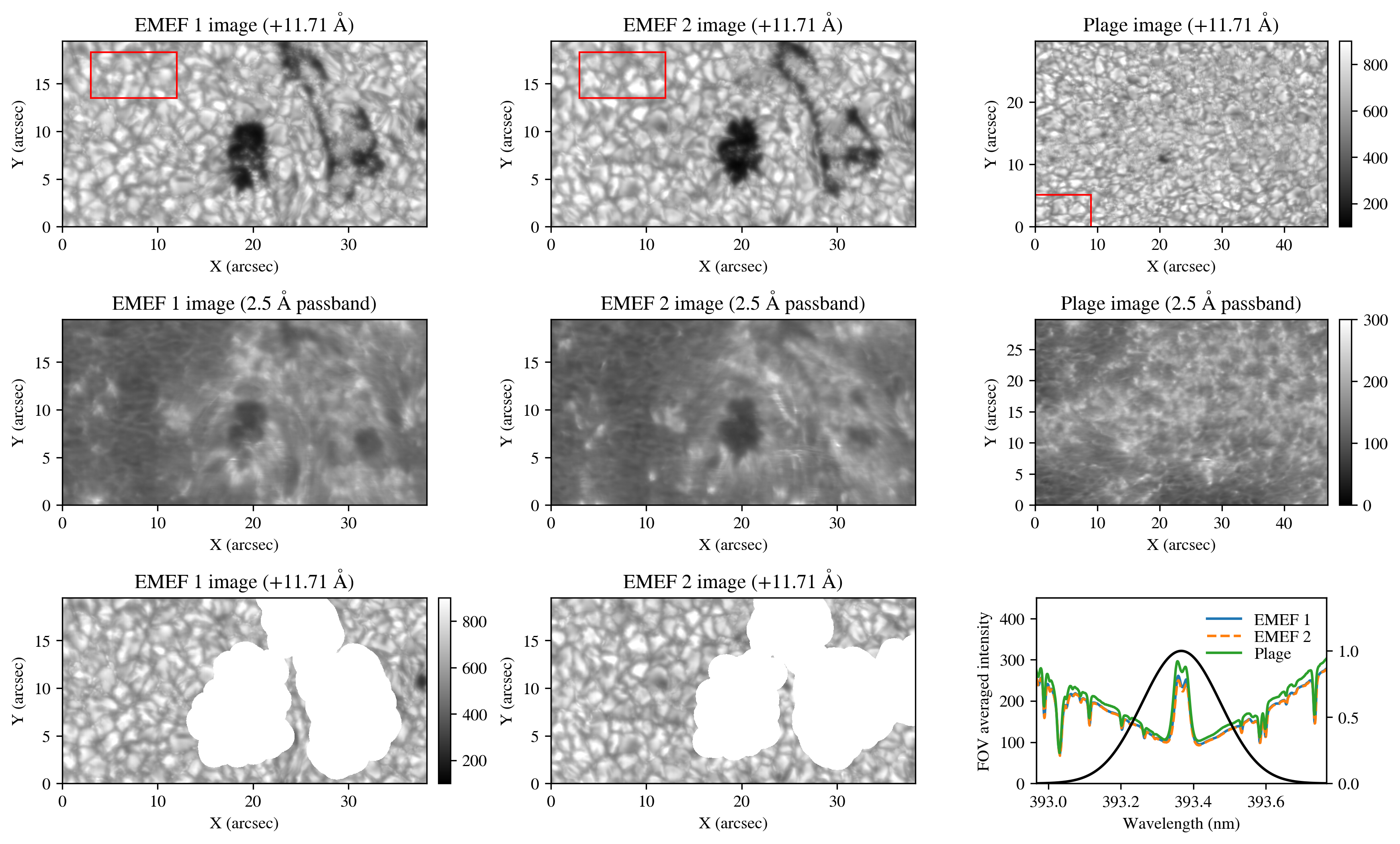}
    \put(4,60){\small\textcolor{black}{\textbf{(a)}}}
    \put(39,60){\small\textcolor{black}{\textbf{(b)}}}
    \put(74,60){\small\textcolor{black}{\textbf{(c)}}}
    \put(4,40){\small\textcolor{black}{\textbf{(d)}}}
    \put(39,40){\small\textcolor{black}{\textbf{(e)}}}
    \put(74,40){\small\textcolor{black}{\textbf{(f)}}}
    \put(4,19.5){\small\textcolor{black}{\textbf{(g)}}}
    \put(39,19.5){\small\textcolor{black}{\textbf{(h)}}}
    \put(74,19.5){\small\textcolor{black}{\textbf{(i)}}}
    \end{overpic}
    \caption{Observed near-continuum images (+11.7~\AA\ from \ca\ line core) of the regions referred to as EMEF~1 (a), EMEF~2  (b), and Plage (c); see main text for more information. Red boxes show the selected QS regions. The grayscale is shown on the right in units of digital numbers. The vertical axis corresponds to the SUSI slit scan direction. The middle row shows spectrally averaged images obtained with a 2.5~\AA\ passband centered at the line core. Panels (g) and (h) show the masks used to exclude spots and pores from the EMEF~1 and 2, respectively. Panel~(i) shows FOV-averaged \ca\ line profiles for the EMEF~1, 2, and Plage, together with a Gaussian profile (black line) of 2.5~\AA\ FWHM. The left y-axis corresponds to the intensity, and the right y-axis to the Gaussian.}
    \label{fig:emef_1_2_plage_cont_2_5_passband_spot_masked}
\end{figure}

\section{Methods}
\label{sec:methods}

{\em Emulation of different passbands.}
To emulate observations obtained with different passbands, we convolved the SUSI spectra with Gaussian profiles, centered at the \ca\ line core.
The full width at half-maximum (FWHM) of each Gaussian corresponds to the desired passband width.
We used Gaussian profiles to approximate instrumental transmission functions, as the exact transmission curves of historical \ca\ filters are generally unknown.
Figure~\ref{fig:emef_1_2_plage_cont_2_5_passband_spot_masked} shows 
the line profile of the EMEF~1, 2, and Plage averaged over the field of view (FOV), along with a Gaussian of FWHM=2.5\,\AA\ (panel i) and the corresponding emulated filtergrams (panels d--f).

{\em Emulation of different spatial resolutions.}
To account for differences in spatial resolution among  observatories, we degraded the SUSI images by convolving them with two-dimensional Gaussians with FWHM matching the target spatial scale. To mitigate edge artifacts introduced by the convolution, each image was trimmed by half the Gaussian FWHM on all sides.

{\em Quiet-Sun normalization.}
To ensure consistent intensity scales across images with different passbands and spatial resolutions (particularly important for ground-based \ca\ observations), we normalized all images by their respective mean QS intensity, $I_\mathrm{QS}$. Throughout the paper, we refer to this normalized intensity as contrast. 
QS regions were selected as areas showing usual granulation in the continuum images and low brightness in both the line core and the broader passbands (red boxes in Figure~\ref{fig:emef_1_2_plage_cont_2_5_passband_spot_masked}, see also Appendix \ref{appendix:qs_region}).
We note that, given the limited FOV, the selected QS regions are not strictly field-free and likely contaminated by weak plage.
This introduces an uncertainty in the normalization level, which should be kept in mind when interpreting the derived contrast–contrast relationships.
To minimize the effect of neighboring pixels due to spatial degradation, we shrunk each QS mask by half the convolution FWHM.

{\em Exclusion of dark features.}
Our focus is on bright plage structures, which show a line profile different from dark features such as sunspots and pores. The Plage observation contains a small pore (at $x=20,y=10$), while EMEF~1 and 2 include both a spot and pores (panels a--c of Figure~\ref{fig:emef_1_2_plage_cont_2_5_passband_spot_masked}).
We applied size and intensity thresholds to mask these features.
This procedure retained the pore at the right edge of EMEF~1 and the one in the Plage dataset. These retained features are not visible in the emulated images with any passband considered here 
and therefore do not affect our analysis.

To exclude spots and pores, we used thresholds on the size and continuum intensity. Regions with $I<0.84I_{\mathrm{QS}}$ and pixel counts greater than 7000 were considered as spots or pores.
The size threshold differentiates between spots or pores and smaller intergranular lanes.
The specific threshold values were determined empirically to ensure all dark features were removed, even at the cost of slightly overmasking.
To eliminate edge effects from spatial degradation, the masks were dilated by half the Gaussian FWHM. 
Although the parameters are somewhat subjective, our goal was a conservative exclusion of dark areas rather than precise segmentation.
To maintain consistent pixel statistics across different spatial resolutions,
all images and masks were trimmed and dilated by the same amount as needed for the poorest spatial resolution used in this study (6\arcsec).

{\em Ground-based data processing.}
 The ground-based Meudon and Rome/PSPT \ca\ filtergrams were processed following \citet{chatzistergos2018,Chatzistergos_analysis_2019} to remove limb darkening, resulting in images in units relative to the QS level.
The images were then compensated for ephemeris to place the solar north pole at the top. 
For direct pixel-by-pixel comparison, the solar disk dimensions in both datasets were matched by resampling the Meudon image to the spatial resolution of the Rome data.
Resampling was performed in the Fourier domain, by cropping high spatial frequencies in the Meudon image such that the inverse transform produced the same apparent solar radius (in pixels) as in the Rome image.
We also compensated for solar rotation between the two observation times  using the routine propagate\_with\_solar\_surface from SunPy \citep{sunpy_community2020, sunpy_version}.

\section{Results} \label{sec:result}

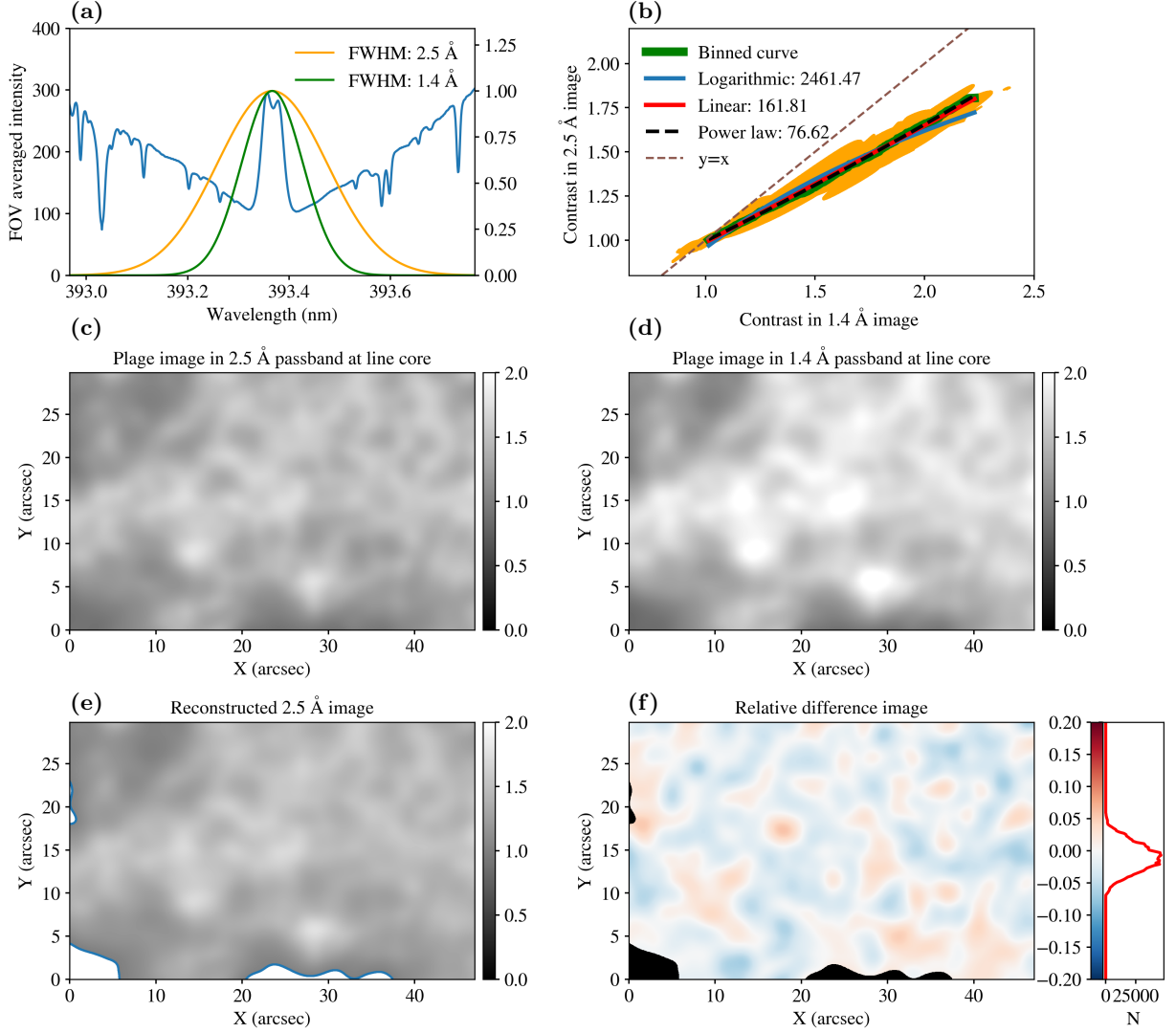
\begin{figure}[ht]
    \centering
    \begin{overpic}[width=0.9\linewidth]{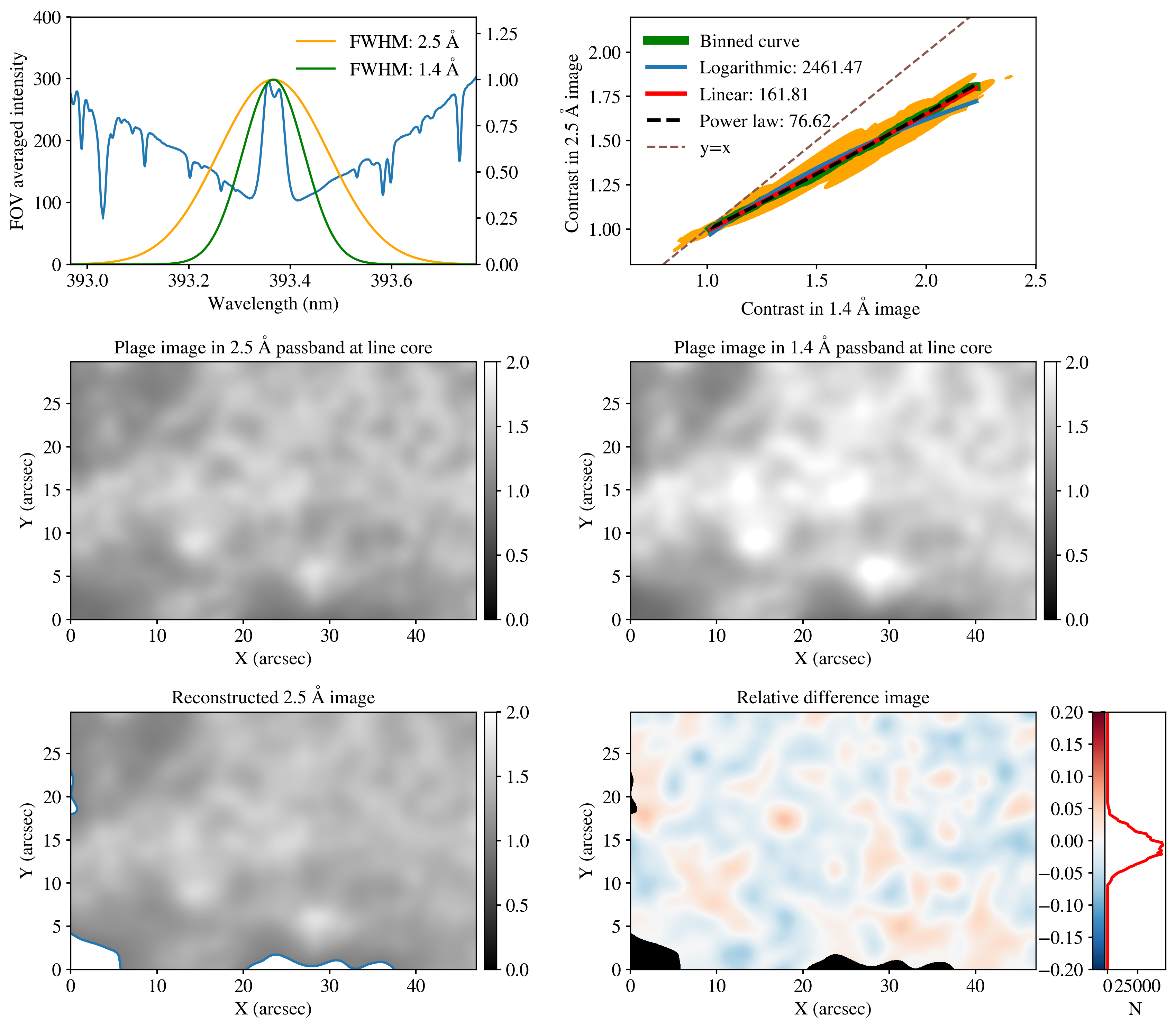}
        \put(6,87){\small\textcolor{black}{\textbf{(a)}}}
        \put(53.5,87){\small\textcolor{black}{\textbf{(b)}}}
        \put(6,60){\small\textcolor{black}{\textbf{(c)}}}
        \put(53.5,60){\small\textcolor{black}{\textbf{(d)}}}
        \put(6,28){\small\textcolor{black}{\textbf{(e)}}}
        \put(53.5,28){\small\textcolor{black}{\textbf{(f)}}}
    \end{overpic}
    \caption{FOV-averaged \ca\ line profile of the Plage region together with the Gaussian profiles of 1.4~\AA\ and 2.5~\AA\ FWHM (panel a).
    The left y-axis corresponds to the intensity, while the right y-axis to the Gaussians. 
    Panel (b) shows the scatter plot of contrast, binned curve, and the fitted linear, logarithmic, and power-law relations, along with the $y=x$ reference line, for pixels from the Plage and EMEF~1 and 2 images in the 1.4~\AA\ and 2.5~\AA\ passbands at spatial resolution of 4\arcsec. The power law exponent is given in Table~1; see main text for details. The reduced $\chi^2$ values for all the fits are provided in their respective legend.  
    The middle row presents the images (normalized by their respective QS mean intensity) in the 2.5~\AA\ and 1.4~\AA\ passbands at a spatial resolution of 4\arcsec. The bottom row displays the normalized 2.5~\AA\ image reconstructed from the 1.4~\AA\ using the derived fitting relation, and the corresponding relative-difference map. Colorbars are shown on the right of each panel. The blue contour in the reconstructed image separates the pixels that are darker (white) and brighter than the $I_\mathrm{QS}$. The same dark region is shown by black in the relative difference map. The histogram next to the colorbar of the relative difference map indicates the pixel count for the relative difference values. 
    }
    \label{fig:1_4_and_2_5_at_2_arcsec_recons_rela_diff}
\end{figure}

\subsection{Deriving the contrast-contrast relationship}
To determine how the observed \ca\ brightness depends on the instrumental passband at a given spatial resolution, we analyzed pairs of emulated images corresponding to different passbands.
For each selected pair and resolution, we generated the corresponding synthetic filtergrams from
all three SUSI datasets (EMEF~1, 2, and Plage) and spatially degraded them to the required resolution. 
The resulting pixel-by-pixel contrast-contrast relationships 
were binned along the x-axis using 3000 data points per bin to reduce scatter.
We adopted the smallest bin size for which the fitted parameters remain stable,
based on sensitivity tests across different passband pairs and spatial resolutions.
These tests showed that increasing the bin size beyond 3000 data points did not lead to significant changes in the parameters.
Although prominent dark features such as sunspots and pores were already excluded, some pixels still exhibit contrasts below the QS level in one or both of the emulated passbands.
These pixels form a scattered population at low contrast and follow a different trend than the bright plage component, introducing a change in slope near zero contrast.
Since the aim of this work is to characterize the bright plage relationship, we restrict the analysis to pixels with contrast above the QS level (hereafter, bright pixels) in both passbands. This selection is applied after binning to avoid introducing asymmetries between the two axes.

Inspection of contrast-contrast relationships for various combinations of passbands and spatial resolutions revealed a clear non-linear dependence.
Therefore, we tested logarithmic and power-law fits and found the best description to be given by the power law of the form $y=ax^b + c$, where $x$ and $y$ are contrasts in the two compared passbands and $a, b, c$ are the coefficients. As an example, Figure~\ref{fig:1_4_and_2_5_at_2_arcsec_recons_rela_diff} shows the relationship between the 1.4~\AA\ and 2.5~\AA\ passbands at a spatial resolution of 4\arcsec.
Panel (b) presents the scatter plot and the fitted functions (linear, logarithmic, and power-law), together with their corresponding reduced $\chi^2$ values (see legend in panel b for specific values).
The power-law fit reproduces the data more accurately than the other functions and yields the lowest reduced $\chi^2$. The coefficients, along with their associated uncertainties, are listed for the power-law fit in the first row of Table~\ref{tab:coefficients}.

To verify the robustness of the derived relationship, we applied it to the Plage dataset to reconstruct an image in one passband from another and then compared the result with the original image. Figure~\ref{fig:1_4_and_2_5_at_2_arcsec_recons_rela_diff} demonstrates this for the case when we reconstructed the 2.5~\AA\ image from the 1.4~\AA\ data at a spatial resolution of 4\arcsec. 
The reconstructed image looks similar to the true 2.5~\AA\ image (Figure \ref{fig:1_4_and_2_5_at_2_arcsec_recons_rela_diff}, panels c--f), with the relative-difference map showing no systematic pattern.
Most pixels differ by less than 7\% (mean difference $\approx2$\%). The root-mean-square (RMS) difference between the original 1.4~\AA\ and 2.5~\AA\ images is 0.2161, whereas that between the reconstructed and original 2.5~\AA\ images is only 0.0342.
This demonstrates that applying the derived power-law relation to convert \ca\ contrast between different passbands improves their consistency, providing a tool for cross-calibrating observations from instruments with diverse bandwidths.

\begin{figure}[!tb]
    \centering
    \begin{overpic}[width=0.96\linewidth]{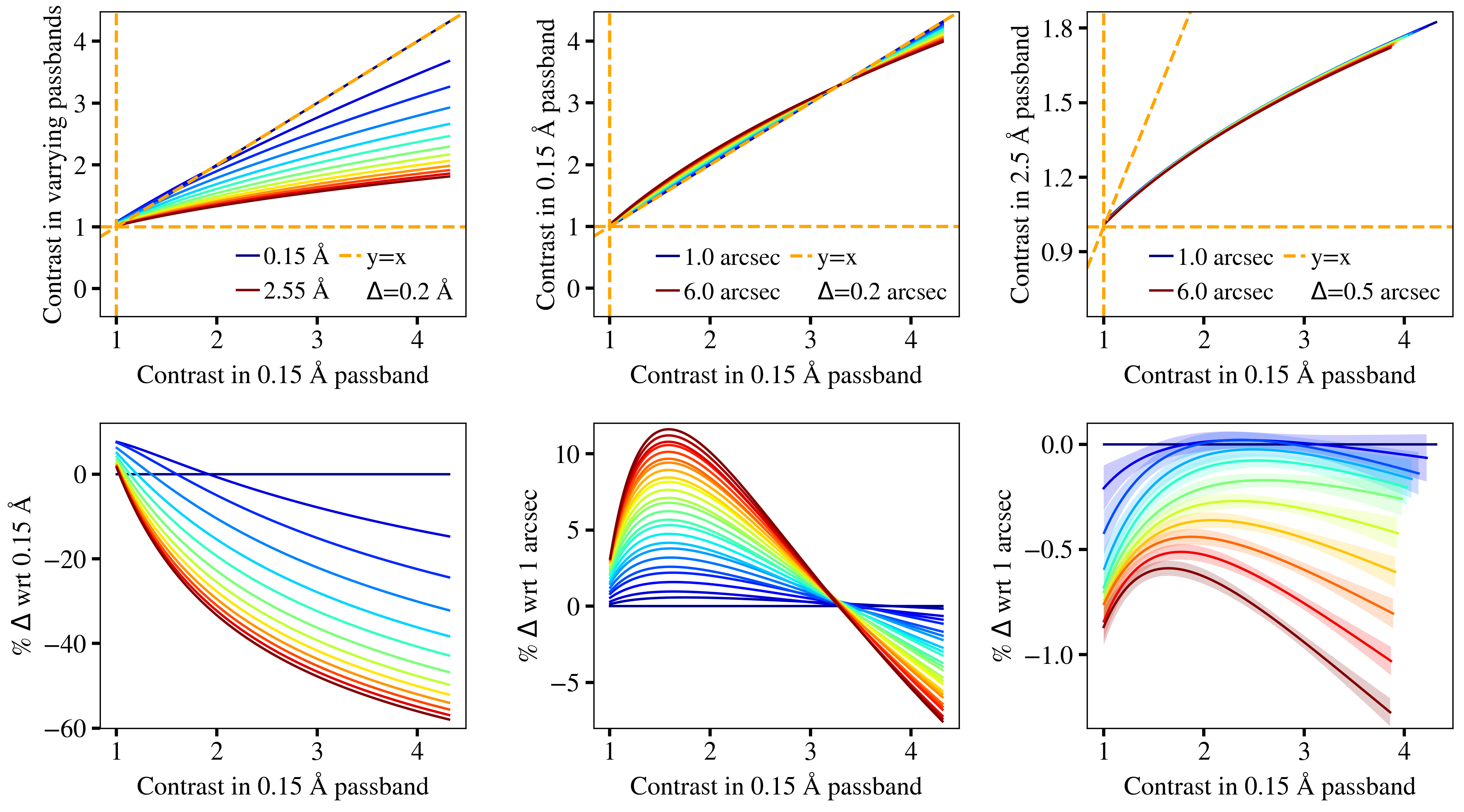}
    \put(17,56){\small\textcolor{black}{\textbf{(a)}}}
    \put(51,56){\small\textcolor{black}{\textbf{(b)}}}
    \put(85,56){\small\textcolor{black}{\textbf{(c)}}}
    \end{overpic}
    \caption{ Top panels: Power-law fits to the contrast-contrast relationships for the following cases: varying passbands at a fixed spatial resolution of 1\arcsec\ (column a); 
    0.15~\AA\ passband at 1\arcsec\ compared with itself at different spatial resolutions (column b); 
    passbands of 0.15~\AA\ and 2.5~\AA\ compared 
    after being degraded to the same spatial resolution, with different curves corresponding to different resolutions (column c).
    Bottom panels: Relative percentage difference of the fits with respect to the reference fit (see Y-axes) vs. contrast in 0.15~\AA\ passband. 
    Passbands in column (a) range from 0.15~\AA\ to 2.55~\AA\ in $\Delta=0.2$~\AA\ steps. Spatial resolutions in columns (b) and (c) range from 1\arcsec\ to 6\arcsec\ in steps of 0.2\arcsec\ and 0.5\arcsec\ for columns (b) and (c), respectively.
    Colors from blue to red indicate increasing passband width or decreasing spatial resolution.
   The dashed orange lines mark $y=x$, $x=1$, and $y=1$. The shaded areas around the fits show the corresponding fit uncertainties.}
    \label{fig:variation_with_spectral_spatial}
\end{figure}

\begin{deluxetable}{ccccccccccc}
\tablecaption{Some example lines of the machine-readable file of the Power-law fit coefficients ($y=ax^b + c$) derived from the \sunriseiii/SUSI observations
\label{tab:coefficients}}
\tablehead{
\colhead{$Px$} & \colhead{$Sx$} & \colhead{$Py$} & \colhead{$Sy$} & \colhead{$a$} & \colhead{$e\_a$} & \colhead{$b$} & \colhead{$e\_b$} & \colhead{$c$} & \colhead{$e\_c$} & \colhead{reduced $\chi ^2$} \\
\colhead{\AA} &\colhead{Arcsec}  & \colhead{\AA} & \colhead{Arcsec} & \colhead{} & \colhead{} & \colhead{} & \colhead{}
}
\startdata
1.4 & 4.0 & 2.5 & 4.0 & 0.4894 & 0.0006 & 1.2322 & 0.0009 & 0.5051 & 0.0006 & 76.6207 \\ 
2.5 & 4.0 & 1.4 & 4.0 & 3.8076 & 0.0118 & 0.4531 & 0.0012 & -2.7988 & 0.01182 & 91.2832 \\
0.15 & 4.0 & 2.5 & 4.0 & 1.0333 & 0.0054 & 0.3939 & 0.0016 & -0.0235 & 0.0055 & 83.8093 \\
\enddata
\tablecomments{Listed are: Passband $Px$, spatial resolution $Sx$, passband $Py$, spatial resolution $Sy$, coefficients $a, b, c$ along with their errors $e\_a, e\_b, e\_c$ and reduced $\chi ^2$ values.}
\end{deluxetable}

\subsection{Effect of passband width and spatial resolution on contrast-contrast relationship}
\label{sec:effect}

The power-law relationships vary with both the passband width and the spatial resolution, see Figure~\ref{fig:variation_with_spectral_spatial}. Column (a) of the figure illustrates the effect of changing the width of one passband while keeping the other fixed at 0.15~\AA\ for a spatial resolution of 1\arcsec. For small differences in bandwidth, the relation remains approximately linear, but it becomes increasingly non-linear as the difference grows. 

Columns (b) and (c) of Figure~\ref{fig:variation_with_spectral_spatial} illustrate how the derived contrast-contrast relationships change with spatial resolution. 
Column (b) shows the effect of degrading spatial resolution for a fixed passband  of 0.15~\AA\  while column (c) examines the case where both passbands (0.15~\AA\ and 2.5~\AA) are degraded to identical spatial resolutions. 
In each column, the fit obtained at the narrowest passband (or highest spatial resolution) is taken as the reference (blue curve). The bottom panels show the relative percentage difference of the other fits with respect to this reference.
The shaded areas around the fits indicate the corresponding fit uncertainties. 

For columns (a) and (b), 
the systematic changes in the fitted relationships clearly exceed the associated uncertainties. 
In contrast, in column (c) the resolution-induced variations are much smaller and are typically comparable to the fit uncertainties.
While these variations are systematic, their amplitude remains well below that produced by changing the passband width, indicating that spatial resolution has a secondary effect on the contrast–contrast relationship when both passbands are degraded together; consistent with the findings of \citet{murabito_investigating_2023}.

The coefficients of the power-law fits for all combinations of passbands (covering 0.1~\AA~to 9~\AA, sampled in steps of 0.05~\AA\ up to 1~\AA\ and 0.1~\AA\ thereafter) and spatial resolutions (1, 2, 4 and 6\arcsec) are provided in the supplementary material, see Table \ref{tab:coefficients} for examples.

\subsection{Application to historical \ca\ data: Converting the Meudon passband to the Rome/PSPT passband}
\label{sec:rome-meudon}

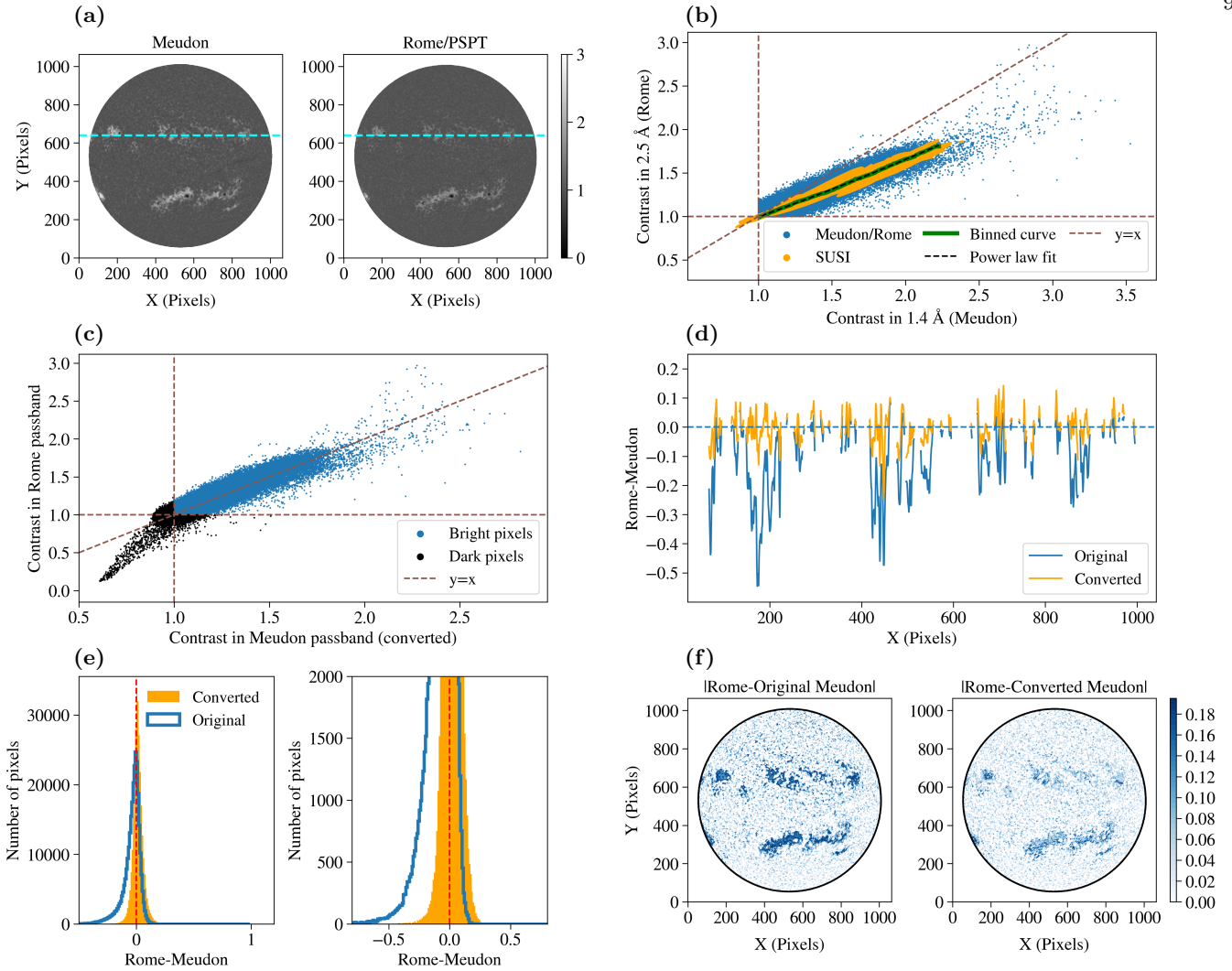
\begin{figure}[ht]
    \centering
    \begin{overpic}[width=0.98\linewidth]{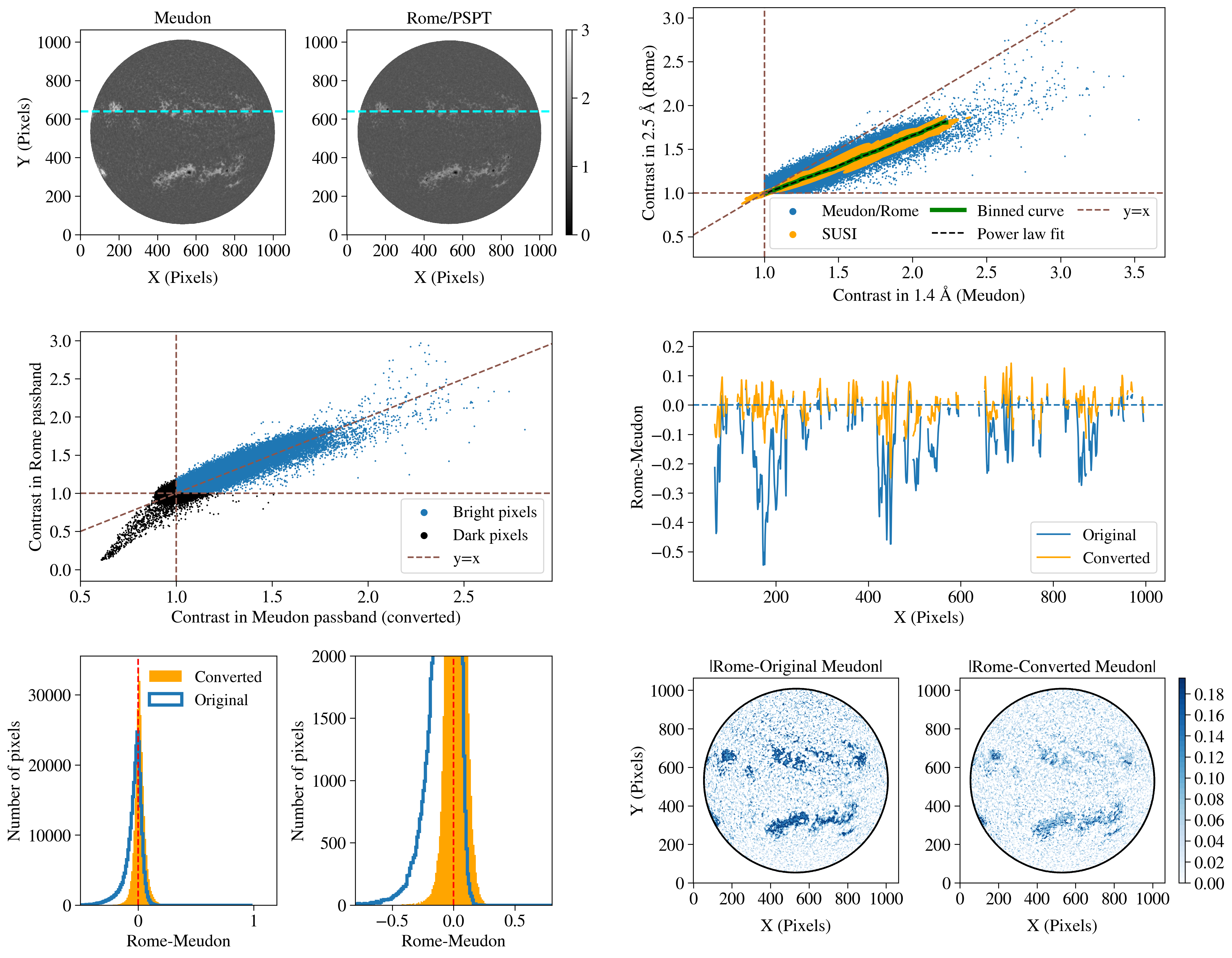}
        \put(6,78){\small\textcolor{black}{\textbf{(a)}}}
        \put(56,78){\small\textcolor{black}{\textbf{(b)}}}
        \put(6,52){\small\textcolor{black}{\textbf{(c)}}}
        \put(56,52){\small\textcolor{black}{\textbf{(d)}}}
        \put(6,25.5){\small\textcolor{black}{\textbf{(e)}}}
        \put(56,25.5){\small\textcolor{black}{\textbf{(f)}}}
    \end{overpic}
\caption{(a): Full-disk \ca\ images from Meudon (1.4~\AA\ passband) and Rome/PSPT (2.5~\AA\ passband) on 1 August 2012.
The grayscale shows the contrast relative to QS.
The dashed cyan lines indicate a horizontal cut across bright regions on the disk. 
(b): Scatter plot of the contrast-contrast relationship between the Meudon and Rome images (blue) compared with the same nominal passbands emulated from SUSI data (orange). The green line shows the binned curve for the emulated data (bin size of 3000 points), and the dashed black line indicates the corresponding power-law fit.
The dashed brown lines mark $y=x$, $x=1$, and $y=1$.
(c): Scatter plot of the contrast-contrast relationship after converting the Meudon image to the Rome passband. 
(d): Difference in contrast between Rome and original (blue) and converted (orange) Meudon across the horizontal cut in panel (a). 
(e): Histogram of the contrast difference between Meudon and Rome images before and after conversion; the right-hand panel shows a zoomed-in view of the distribution. 
(f): Full-disk maps of the absolute difference between Rome and Meudon before and after conversion, with contours corresponding to different levels of absolute difference.}
    \label{fig:rome_meudon}
\end{figure}

To test whether the relationship derived from SUSI data improves the cross-calibration of other \ca\ observations, we applied it to Meudon (degraded to the spatial resolution of Rome/PSPT) and Rome full-disk filtergrams, converting the Meudon (narrower  1.4~\AA\ passband) image to the equivalent of the Rome (broader 2.5~\AA\ passband) image. The contrast of the two datasets clearly differs, with the Meudon image showing higher contrast than the Rome image  (Figure~\ref{fig:rome_meudon}a), consistent with Meudon's narrower bandwidth.
The scatter plot of their pixel-by-pixel contrast (Figure~\ref{fig:rome_meudon}b) follows the same non-linear trend found in the SUSI-based emulations, albeit with greater dispersion.
The latter is due to factors such as time difference between observations, variable seeing, uncertainties in our spatial resolution estimates, and possible deviations between nominal and actual instrumental passbands.
The close agreement confirms that, in spite of the above-mentioned uncertainties, the observed contrast offset primarily arises from the difference in passbands, allowing the SUSI-derived power-law to be used for inter-calibration. The first row of Table \ref{tab:coefficients} shows the fit coefficients. As the contrast-contrast relationship has been derived using only bright pixels (contrast higher than QS), we limit the application of the relationship to only these bright pixels. After applying the calibration to convert the Meudon data to a Rome-like passband, the scatter distribution concentrates around the $y=x$ line (Figure~\ref{fig:rome_meudon}c), demonstrating a significant improvement in consistency between the two datasets. 
For reference, we also show the application of the power fit over the dark pixels (of sunspots, pores, and those darker than QS contrast; see Figure~\ref{fig:rome_meudon}c).

To quantify the conversion performance, we compared contrast differences along a representative  horizontal cut across the solar disk. The cut is shown as dashed cyan lines in Figure~\ref{fig:rome_meudon}a, and the difference across the cut is presented in Figure~\ref{fig:rome_meudon}d. After the conversion, the difference between the Meudon and Rome contrast decreased significantly, with the RMS difference along the cut reducing from 0.1682 to 0.0514. More importantly, there is no systematic offset between the contrasts in the two datasets. 
A histogram of pixel-wise differences over the full disk (Figure~\ref{fig:rome_meudon}e) confirms this improvement.
The original histogram is skewed towards negative values while after conversion it centers around zero.
The full disk RMS difference decreased from 0.0940 to 0.0492.

Since the SUSI data used to derive the contrast-contrast relationship were recorded near the disk center, applying it to full-disk conversion implicitly assumes that there is no significant variation in the relationship across the solar disk.
To test this assumption, we examined maps of the absolute difference between the Meudon and Rome images before and after conversion (Figure~\ref{fig:rome_meudon}f).
No systematic center-to-limb variation (CLV) was found, confirming that the conversion remains valid across the disk (see also Appendix \ref{appendix:clv}). Our results show that the power-law relation derived from SUSI data significantly improves cross-calibration between Meudon and Rome images obtained with different passbands. 

We also investigated the dependence of the conversion performance on the direction of the conversion and found it to be asymmetric. 
Converting a narrow-band to a broader passband works better than the reverse (compare Figure~\ref{fig:rome_meudon} with Figure~\ref{fig:meudon_calibrated_rome} in the Appendix \ref{appendix:cali_broad_narrow}). 
We believe that asymmetry arises because a narrow passband samples a thinner, higher atmospheric layer, preserving more information on bright plage pixels, whereas broader passbands average their contributions over a thicker layer, effectively diluting them. The brightness sensed in very narrow bands is also far more sensitive to Doppler shifts of the line than in broader bands.
Converting a narrow-band image into a broad-band one is equivalent to reducing the contrast of plage pixels, which the power-law fit does well.
The opposite conversion is less effective because information lost
in the broader passband cannot be recovered statistically.
Hence, whenever possible, cross-calibration of \ca\ archives should be performed from narrower to broader passbands.

Finally, although the difference between the Meudon (1.4~\AA) and Rome (2.5~\AA) passbands is moderate, we have verified that the method also performs well for passbands with a larger width mismatch (see Appendix \ref{appendix:high_passband_diff}).
This additional test confirms that the conversion remains effective even in cases of substantial passband disparity.

\section{Discussion and Conclusions}
\label{sec:discussion and Conclusions}

In this study, we quantified how differences in instrumental passbands and spatial resolutions affect \ca\ observations.
Using high-spectral-resolution data from the SUSI instrument, we derived a power-law relation that enables conversion of \ca\ contrast between observations obtained with different passbands.
This approach provides a means to homogenize data from diverse historical and modern archives. The derived power-law relation successfully accounts for the differences between passbands, allowing accurate conversion of images taken in different passbands.
Applying this relation for cross-calibration brings the corresponding images into close agreement.

One of the main challenges is determining a representative QS level within the limited FOV of the available datasets.
The region used for normalization in the SUSI data likely corresponds to weak plage rather than perfectly QS.
This introduces a systematic uncertainty in the absolute contrast scale.
Because the same normalization is applied consistently to both passbands, the contrast–contrast relationship is modified along both axes, although not necessarily by the same factor. 
Consequently, the absolute scaling of the relation is affected by the adopted QS level, while the overall mapping between passbands remains similar.
The performance of the derived relationship is supported by its ability to reproduce the observed contrasts when converting between passbands.
Future studies could improve the derived relationship by determining $I_\mathrm{QS}$ from a dedicated SUSI QS dataset.
A more accurate determination of the QS level in archival \ca\ observations would likewise improve the quantitative calibration of those data.

Although the derived relation performs well for the studied
passband pair, applying it to data from other observatories requires care.
The actual passbands and spatial resolutions of ground-based observations are often uncertain.
In this study, we assumed ideal Gaussian passbands and a simple scaling between spatial resolution and pixel scale.
In practice, passbands may differ from their nominal specifications (for example, due to tilt in the filter or temperature changes), and seeing conditions cause time-dependent variations in spatial resolution \citep{Ermolli_spatial_res}.
Future applications should therefore consider the true transmission profiles and spatial resolutions of each observation.

Despite these limitations, the high and constant spectral and spatial resolution of SUSI observations allows us to emulate different passbands and spatial resolutions found in historical \ca\ archives.
The derived power-law relation describes the contrast-contrast relationship between different passbands well and shows, within the range studied here, a much stronger sensitivity to passband width than to spatial resolution. 
Applying this relation to independent ground-based data confirms that it indeed enables accurate conversion between passbands.
Further, we found no significant CLV effect on the conversion.

These results provide a tool for the cross-calibration of \ca\ images from different observatories, laying the foundation for the creation of a long, homogeneous time series of full-disk plage and magnetic-field records.
Such a dataset will allow improved solar irradiance reconstruction based directly on facular information, help resolve discrepancies among existing models, and offer valuable insights into the long-term evolution of the solar magnetic field and dynamo.


\clearpage
\newpage
\begin{acknowledgments}
AKY acknowledges support within the framework of the International Max-Planck Research School (IMPRS) for Solar System Science at the Technical University of Braunschweig.
FAI is member of the ``Carrera del Investigador Cient\'ifico" of CONICET.
 \sunriseiii\ is supported by funding from the Max-Planck-Förderstiftung (Max Planck Foundation), NASA under Grants \#80NSSC18K0934 and \#80NSSC24M0024 (“Heliophysics Low Cost Access to Space” program), and the ISAS/JAXA Small Mission-of-Opportunity program and JSPS KAKENHI Grant Numbers JP18H05234 and JP23K25916. This research has received financial support from the European Union’s Horizon 2020 research and innovation program under grant agreement No. 824135 (SOLARNET) and No. 101097844 (WINSUN) from the European Research Council (ERC). It has also been funded by the Deutsches Zentrum für Luft- und Raumfahrt e.V. (DLR, grant no. 50 OO 1608). The Spanish contributions have been funded by the Spanish MCIN/AEI under projects RTI2018-096886-B-C5, and PID2021-125325OB-C5, and from “Center of Excellence Severo Ochoa” awards to IAA-CSIC (SEV-2017-0709, CEX2021-001131-S), all co-funded by European REDEF funds, “A way of making Europe".
This research has made use of the Astrophysics Data System  (ADS; \url{https://ui.adsabs.harvard. edu/}) Bibliographic Services, funded by NASA under Cooperative Agreement 80NSSC21M00561.

\end{acknowledgments}
\software{SunPy \citep{sunpy_community2020, sunpy_version}}

\appendix

\section{Selection of QS region}
\label{appendix:qs_region}

\begin{figure}[ht]
    \centering
    \includegraphics[width=1\linewidth]{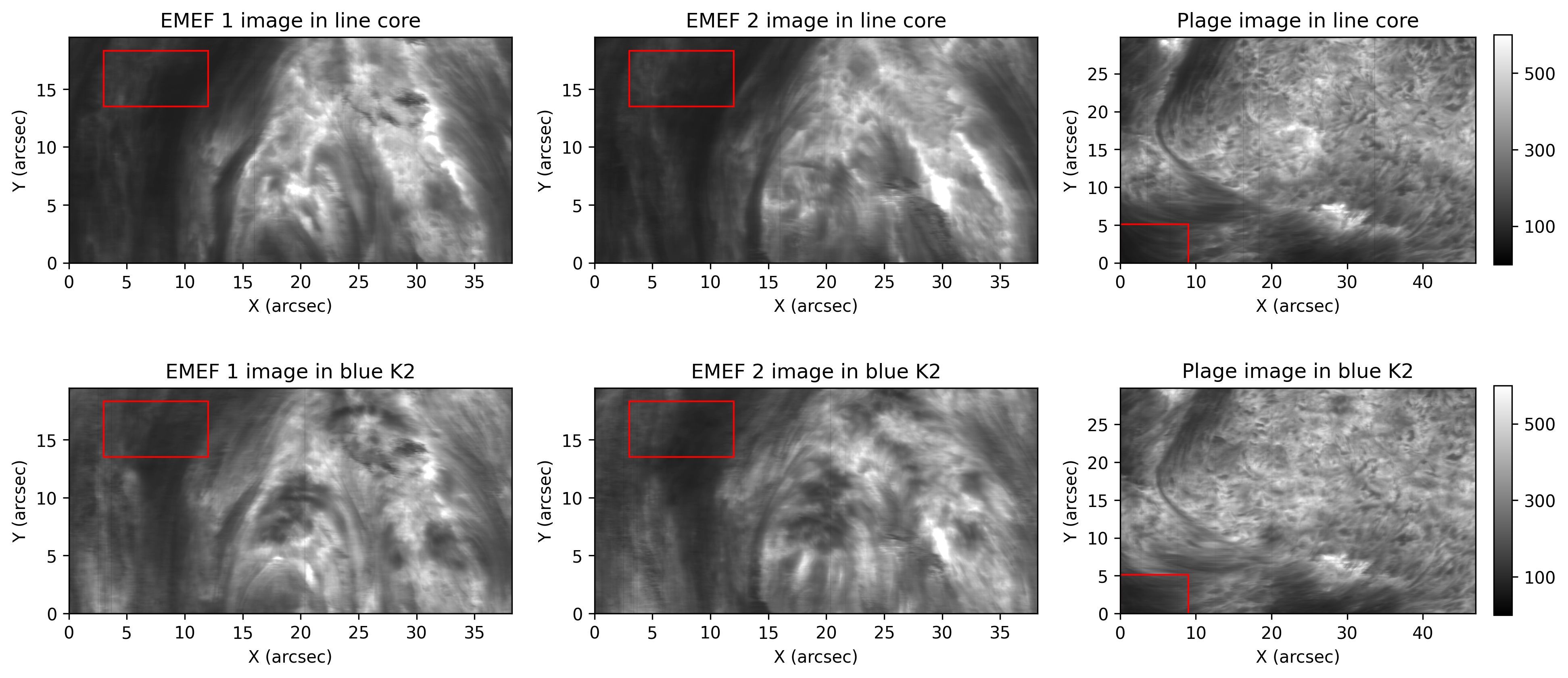}
    \caption{EMEF~1, 2, and Plage observations in \ca\ line core and blue K2. Red boxes show the selected QS regions. The grayscale is shown on the right in units of digital numbers. The vertical axis corresponds to the SUSI slit scan direction.}
    \label{fig:line_core_qs}
\end{figure}

The limited field of view of the SUSI observations makes the identification of representative quiet-Sun (QS) regions non-trivial. 
In addition, because our analysis involves multiple passbands, the selected QS regions must remain quiet when viewed through all passbands considered.
To ensure this, QS regions were identified in the continuum images as areas exhibiting normal granulation and low magnetic activity, see panels a--c of Figure~\ref{fig:emef_1_2_plage_cont_2_5_passband_spot_masked}. Further they were required to appear dark in the \ca\ line-core and K2 observations. Consequently, regions that are quiet in the line core and K2 remain quiet when observed with broader passbands.
This criterion guarantees a consistent QS reference across all emulated passbands.
Figure \ref{fig:line_core_qs} illustrates the selected QS regions in the \ca\ line-core and K2.

\section{Center-to-Limb Variation in the performance of conversion}
\label{appendix:clv}
\begin{figure}[ht]
    \centering
    \includegraphics[width=0.6\linewidth]{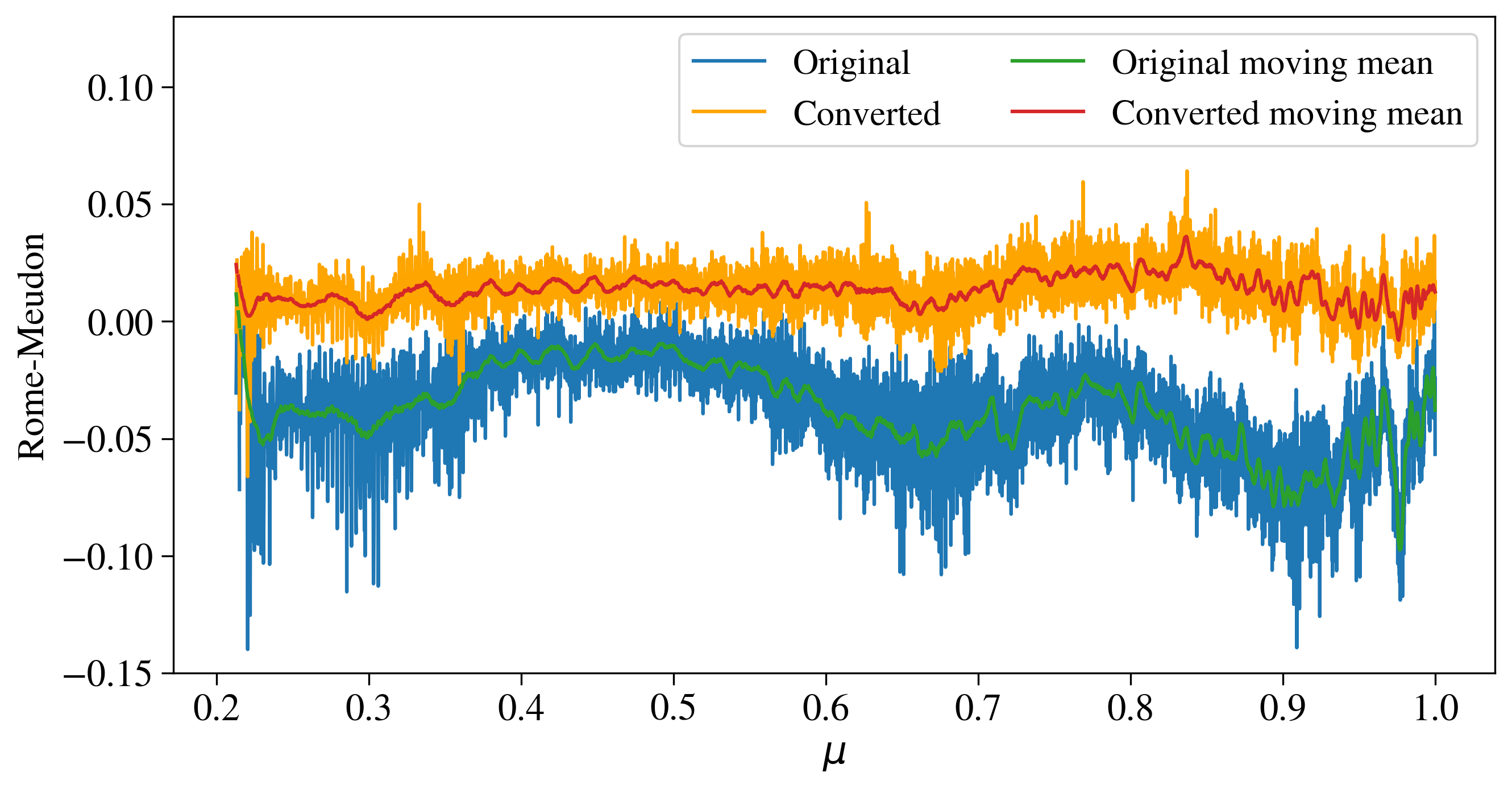}
    \caption{CLV of the difference between the Rome image and the Meudon images before and after conversion of the Meudon image to the Rome image. The moving mean has a window size of 100 points.}
    \label{fig:clv}
\end{figure}

Full-disk absolute difference images (Figure~\ref{fig:rome_meudon}f) show no clear CLV effect on conversion performance.
To test this quantitatively, we sorted pixels by their distance from the disk center and grouped them into blocks of 100.
For each block, we computed the mean difference and the mean value of $\mu$ (the cosine of the heliocentric angle).
Figure~\ref{fig:clv} shows the contrast difference between pixels as a function of $\mu$ before and after conversion.
Although some residual CLV trend is present, it is weak, suggesting that the conversion remains nearly uniform across the solar disk.

\section{Converting a broad passband to a narrow passband}
\label{appendix:cali_broad_narrow}
\begin{figure}[ht]
    \centering
    \begin{overpic}[width=1\linewidth]{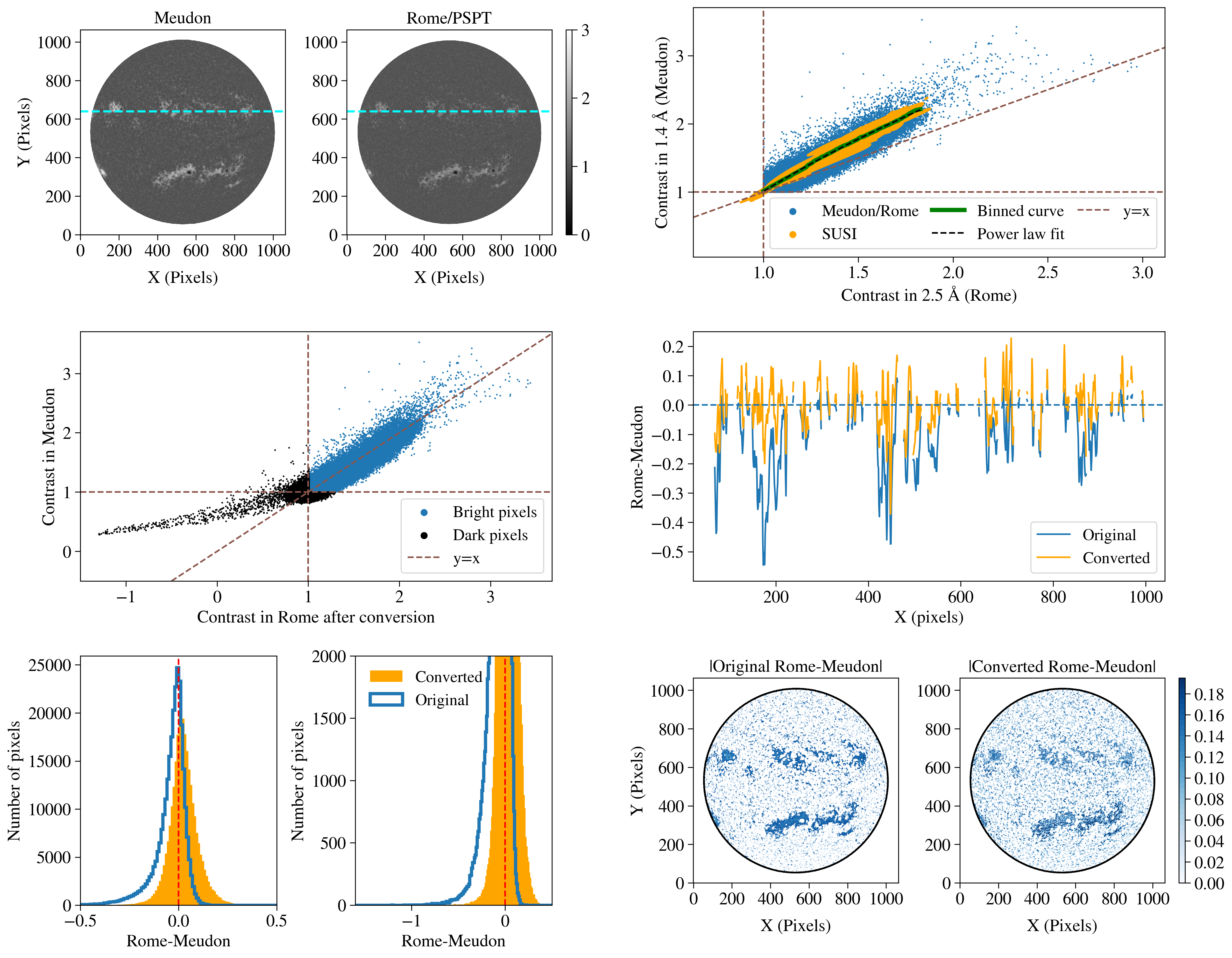}
        \put(6,78){\small\textcolor{black}{\textbf{(a)}}}
        \put(56,78){\small\textcolor{black}{\textbf{(b)}}}
        \put(6,52){\small\textcolor{black}{\textbf{(c)}}}
        \put(56,52){\small\textcolor{black}{\textbf{(d)}}}
        \put(6,25.5){\small\textcolor{black}{\textbf{(e)}}}
        \put(56,25.5){\small\textcolor{black}{\textbf{(f)}}}
    \end{overpic}
\caption{Same as Figure~\ref{fig:rome_meudon} but now broader passband (2.5~\AA) image is converted to a narrower passband (1.4~\AA) image.}
\label{fig:meudon_calibrated_rome}
\end{figure}

In Section \ref{sec:rome-meudon}, we converted the Meudon (narrow passband) image to the Rome (broad passband) image and found good agreement.
Here, we perform the reverse test, converting the Rome (broad) image to the Meudon (narrow) image. Figure~\ref{fig:meudon_calibrated_rome} shows the results, and Table~\ref{tab:coefficients} (second row) lists the fit coefficients. Panels d--f clearly demonstrate that conversion in this direction is notably less effective than in the narrow-to-broad case (compare with Figure~\ref{fig:rome_meudon}).
The RMS difference before and after conversion is 0.1682 and 0.0807 across the horizontal cut, and 0.0940 and 0.0779 over the full disk, respectively. 
We note that for dark pixels, the contrast-contrast relationship yields negative contrast values, which are not physical (panel c of Figure~\ref{fig:meudon_calibrated_rome}). 

\section{Conversion of the Meudon spectroheliograph passband to the Rome/PSPT passband}\label{appendix:high_passband_diff}

\begin{figure}[ht]
    \centering
    \begin{overpic}[width=1\linewidth]{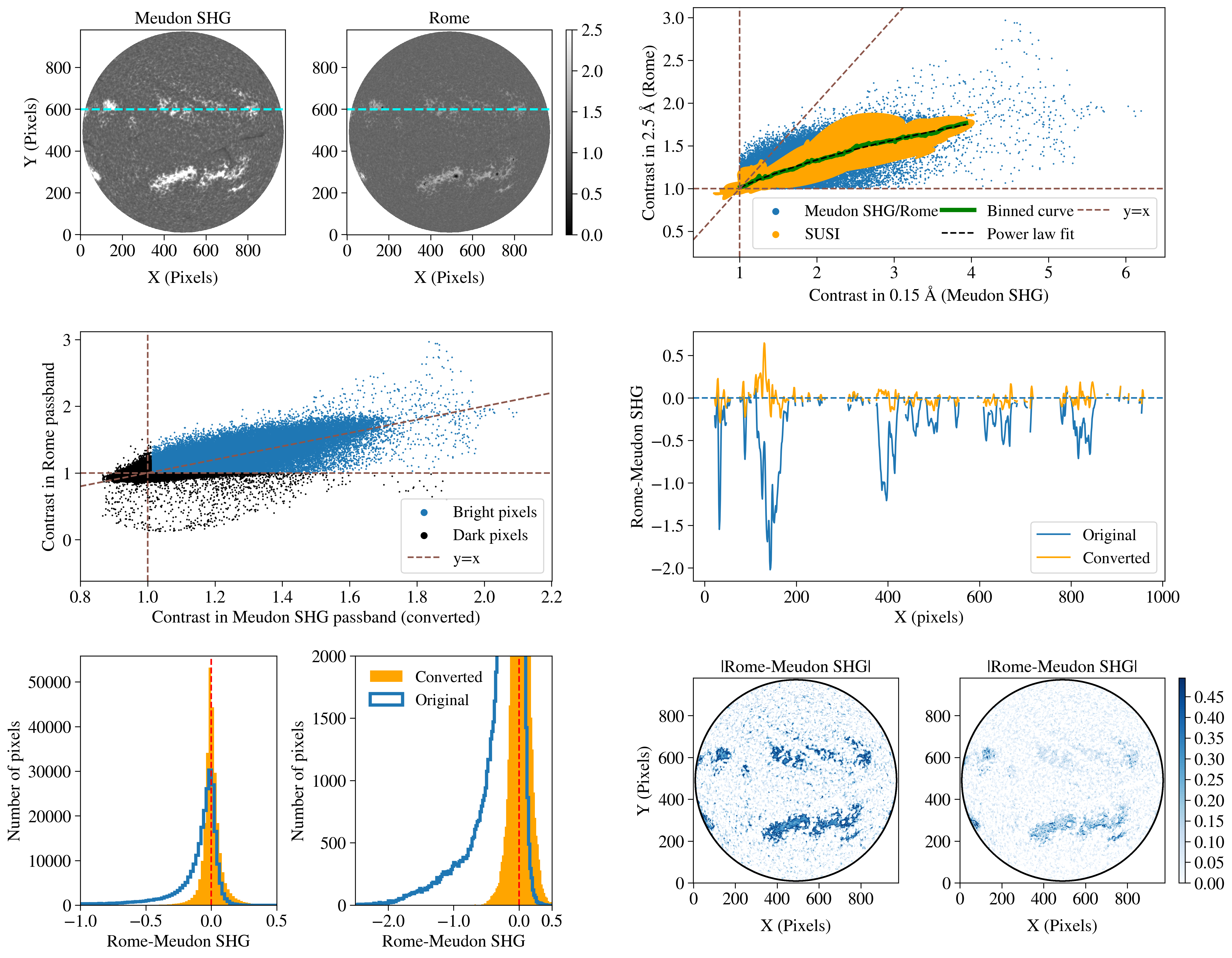}
        \put(6,78){\small\textcolor{black}{\textbf{(a)}}}
        \put(56,78){\small\textcolor{black}{\textbf{(b)}}}
        \put(6,52){\small\textcolor{black}{\textbf{(c)}}}
        \put(56,52){\small\textcolor{black}{\textbf{(d)}}}
        \put(6,25.5){\small\textcolor{black}{\textbf{(e)}}}
        \put(56,25.5){\small\textcolor{black}{\textbf{(f)}}}
    \end{overpic}
\caption{Same as Figure~\ref{fig:rome_meudon} but for Meudon SGH passband (0.15~\AA) image and Rome passband (2.5~\AA) image.}
\label{fig:meudon_shg_rome}
\end{figure}
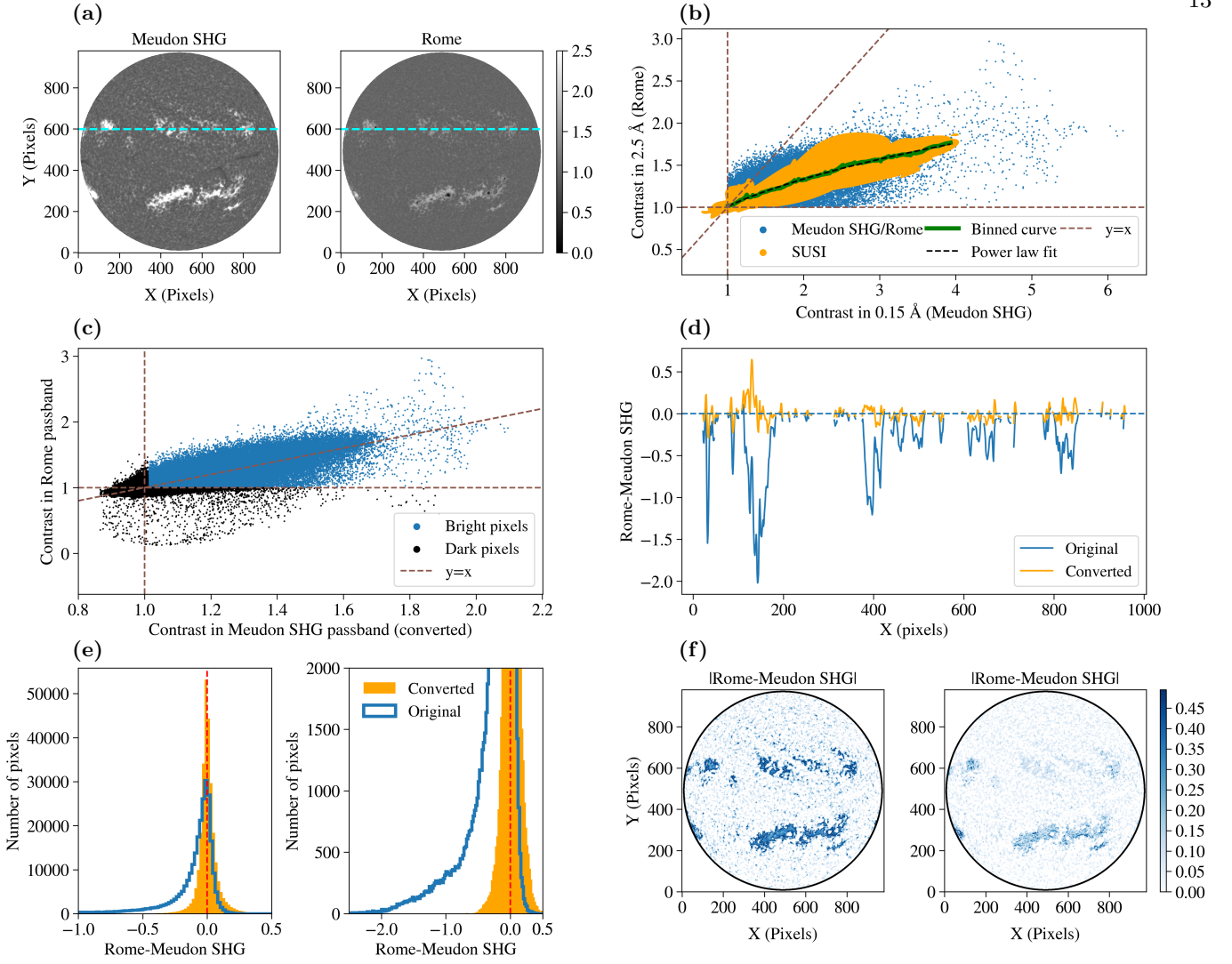

To test conversion performance for more extreme differences in passband width, we compared the Rome/PSPT observation described in Section~\ref{sec:data} with an observation from Meudon spectroheliograph \citep[][]{malherbe_130_2023}.
We refer to this observation as Meudon SHG (for Spectro-Helio-Gram) to distinguish it from the Meudon heliograph observation used in Section~\ref{sec:rome-meudon}.
The heliograph produces an image using a filter, whereas the spectroheliograph builds an image by scanning the solar disk with a slit. The Meudon SHG observation was taken on 1~August 2012 at 07:25:24~UT with spectral passband of 0.15~\AA\ and pixel scale of 1.5\arcsec~$\text{pixel}^{-1}$. 
The Rome observation was taken on the same day at 07:08:37~UT.
To account for spatial resolution and the difference in the time of observation, we applied the same procedures as used for the Meudon--Rome comparison. 
The results are shown in Figure~\ref{fig:meudon_shg_rome}.
They closely mirror those for Meudon--Rome: the Meudon SHG image has higher contrast values than the Rome image, the scatter plot of the contrast-contrast relation matches that derived from the SUSI data, and after conversion the points align well along the $y=x$ line. 
The coefficients of the power-law fit, along with the reduced $\chi^2$ value, are listed in the third row of Table~\ref{tab:coefficients}.
The contrast difference along a representative horizontal cut and over the full disk is significantly reduced after conversion, with RMS differences of 0.5158 and 0.1022 across the cut, and 0.3193 and 0.0764 over the full disk, before and after conversion, respectively.
No spatial non-uniformity or CLV is evident.
These results demonstrate that the power-law relation derived from SUSI performs reliably even for large differences in passband width.


\newpage

\clearpage
\newpage
\bibliography{sample701}{}
\bibliographystyle{aasjournalv7}



\end{document}